\documentclass[reprint,
amsmath,amssymb,
prl, 
]{revtex4-1}
\usepackage{graphicx}
\usepackage{dcolumn}
\usepackage{bm}

\usepackage{placeins}

\newcommand{\x}[0]{\mathbf{x}}
\newcommand{\dx}[0]{\mathrm{d}\x}
\DeclareMathOperator{\EX}{\mathbb{E}}
\newcommand{\Ha}[0]{H_A(\x)}
\newcommand{\Hb}[0]{H_B(\x)}

\begin{document}

	\title{Small Sample Limit of the Bennett Acceptance Ratio Method and the Variationally Derived Intermediates}
	
	\author{Martin Reinhardt}

	\author{Helmut Grubm\"uller}%
	\email{hgrubmu@gwdg.de}
	\affiliation{%
		Max Planck Institute for Biophysical Chemistry, Am Fassberg 11, 37077 G\"ottingen, Germany
	}%

	\date{\today}

	\begin{abstract}
		Free energy calculations based on atomistic Hamiltonians provide microscopic insight into the thermodynamic driving forces of  biophysical or condensed matter systems. Many approaches use intermediate Hamiltonians interpolating between the two states for which the free energy difference is calculated. The Bennett Acceptance Ratio (BAR) and Variationally Derived Intermediates (VI) methods are optimal estimator and intermediate states in that the mean-squared error of free energy calculations based on independent sampling is minimized. However, BAR and VI have been derived based on several approximations that do not hold for very few sample points. Analyzing one-dimensional test systems we show that in such cases BAR and VI are suboptimal and that established uncertainty estimates are inaccurate. Whereas for VI to become optimal less than seven samples per state suffice in all cases, for BAR the required number increases unboundedly with decreasing configuration space densities overlap of the end states. We show that for BAR the required number of samples is related to the overlap through an inverse power law. Because this relation seems to hold universally and almost independent of other system properties, these findings can guide the proper choice of estimator for free energy calculations.
	\end{abstract}
	
	\maketitle
	
	\section{Introduction}
	
    Free energy differences provide detailed insights into the molecular driving forces of biophysical processes and their accurate calculation is crucial for their successful application, e.g., in pharmaceutical ligand design or material science~\cite{Chipot2007, Ge2016, Sun2018, Cournia2017, Swinburne2018, Baumeler2019, Armacost2020}. To calculate the free energy difference between, e.g., two potential drug molecules bound to a receptor, alchemical equilibrium techniques~\cite{Kirkwood1935} based on simulations with atomistic Hamiltonians are amongst the most widely used methods. Aside from the two states of interest, these techniques conduct sampling from intermediate states whose Hamiltonians are constructed from those of the end states. The step-wise summation of the individual differences then yields the total free energy difference.
    
    Two choices have to be made that critically affect the accuracy of free energy calculations: 
    Firstly, the choice of the estimator that is used to evaluate the free energy differences between the individual states. Whereas a number of estimators exist that have practical advantages in different situations~\cite{Zwanzig1954, Kirkwood1935, Wu2005}, it has been shown that between two states the Bennett Acceptance Ratio (BAR) method \cite{Bennett1976} minimizes not only the variance, but also the mean-squared error (MSE) \cite{Reinhardt2020}. Remarkably, as will be revisited in the theory section, the Zwanzig formula~\cite{Zwanzig1954} yields identical MSEs if applied together with an optimally chosen virtual intermediate state in which no sampling is conducted~\cite{Wu2005, Reinhardt2020}. For BAR, the variance and the bias have been extensively analyzed~\cite{Shirts2005, Wu2005, Wu2005a, Hahn2009, Konig2011, Schultz2021}. As the MSE can be decomposed into variance plus the squared bias, and therefore accounts for both the variance and the bias, we will focus our analysis in this work on the MSE. Further, from an application perspective, the MSE is the relevant quantity.
    
    The second choice concerns the functional form of the intermediate states, i.e., how these are constructed from the two end state Hamiltonians. Apart from the conventionally used linear interpolation intermediates, various functional forms have been suggested~\cite{Blondel2004, Christ2007, Perthold2018, Konig2020}, with a particular focus on appearing or vanishing particles in solution~\cite{Steinbrecher2007, Pham2011, Pham2012, Buelens2012, Gapsys2012}. In general, when using the Zwanzig formula or BAR as an estimator, and assuming independent samples, the Variationally-derived Intermediates (VI) \cite{Reinhardt2020, Reinhardt2020a, Reinhardt2021} have been shown to yield the optimal MSE amongst all possible functional forms of intermediate states. 
    
    However, both BAR and VI have been derived using approximations that strictly hold only for large sample numbers. This question becomes particularly urgent for free energy calculations of large systems or when using quantum mechanics based methods~\cite{Beierlein2011, Giese2019, Zhang2019, Hall2020}, which are computationally demanding and, therefore, provide limited sampling. Further, sample points derived from atomistic simulations are time-correlated, such that the effective number of independent sample points is often orders of magnitude smaller than the number of configurations obtained from a simulation. We therefore will analyze how the accuracy of BAR and VI depends on sample size, and show how the obtained scalings provide guidance on their proper use.
    
    \section{Theory}
   
   Several different derivations of BAR have been published \cite{Bennett1976, Shirts2003, Habeck2012}, resting on different assumptions. Here we recapitulate the one with the least restrictive assumptions, which also highlights the unexpected relation between estimators and intermediate states~\cite{Reinhardt2020}. The generalization of this relation to $N$ intermediate states has been used to derive VI. Both approaches 
   rest on the Zwanzig formula~\cite{Zwanzig1954}. Accordingly, the free energy difference between states $A$ and $B$ with Hamiltonians $H_A(\x)$ and $H_B(\x)$, respectively, is given by
   \begin{align}
   \Delta G_{A, B}= -\ln\langle e^{-[H_B(\x) - H_A(\x)]}\rangle_A\;\;,
   \label{eq:zwanzig}
   \end{align}
   where $\x\in {\rm I\!R}^{3M}$ denotes the position of all $M$ particles of the simulation system. Only sample points from state $A$ are used, where $\langle\rangle_A$ denotes the ensemble average. For ease of notation, all energies are expressed in units of $k_BT$.
   
   In the following, the free energy estimate governed by Hamiltonian $H_A(\x)$ that is obtained when the ensemble average in Eq.~(\ref{eq:zwanzig}) is calculated from a finite sample of size $n$ will be denoted by $\Delta G_{A\rightarrow B}^{(n)}$, whereas $\Delta G_{A,B}$ denotes the exact free energy difference. For statistically independent samples, the MSE of the free energy calculated via Eq.~(\ref{eq:zwanzig}) reads~\cite{Reinhardt2020}

\begin{align}
\mathrm{MSE}\left(\Delta G_{A\rightarrow B}^{(n)}\right) =& \EX\left[ \left(\Delta G_{A, B} - \Delta G_{A\rightarrow B}^{(n)} \right)^2\right]\; \label{eq:mse_ansatz}\\  =& \frac{1}{n}\left(\int \frac{\left(p_B(\x)\right)^2}{p_A(\x)}\,\dx\,  - 1\right) \,,
\label{eq:mse_zwanzig}
\end{align}
where $p_A(\x)=e^{-H_A(\x)}/Z_A$ and $p_B(\x)=e^{-H_B(\x)}/Z_B$ denote the configuration space densities and $Z_A$ and $Z_B$ the partition functions of the respective end states.

Importantly, the derivation of the MSE of the Zwanzig formula, Eq.~(\ref{eq:mse_zwanzig}), and therefore also the optimization thereof leading to BAR and VI, is based on approximations. As a prior step, we consider the Hamiltonian $H_B(\x) - C$, i.e., the Hamiltonian of end state $B$ shifted by a constant $C$. Using this Hamiltonian with the Zwanzig formula, Eq.~(\ref{eq:zwanzig}), the free energy difference between $A$ and $B$ is calculated as
   \begin{align}
	\Delta G_{A, B}= -\ln\langle e^{-[H_B(\x) -C - H_A(\x)]}\rangle_A + C\;\;.
	\label{eq:zwanzig_C}
\end{align}
We now denote the sample based average from Eq.~(\ref{eq:zwanzig_C}) as
\begin{align}
y^{(n)}(C) = \frac{1}{n} \sum_{i=1}^{n} e^{-[H_{B}(\x_i) -C- H_A(\x_i)]}
\end{align}
and the exact ensemble average as
\begin{align}
y(C) = \int p_A(\x)\dx\, e^{-[H_{B}(\x) - C - H_A(\x)]}\;\;.
\end{align}

For large $n$, using $C\approx \Delta G_{A,B}$ implies $y^{(n)}(C) \approx y(C) \approx 1$. After expanding the MSE, Eq.~(\ref{eq:mse_ansatz}) (for the full derivation see Ref.~\cite{Reinhardt2020}), the expectation value of the estimate based on finite sampling  
\begin{align}
	\begin{split}
		\EX\left[\Delta G_{A \rightarrow B}^{(n)}\right] &=\\ - \int p_A(\x_1)\dx_1 &... \int p_A(\x_n)\dx_n \ln \left( y^{(n)}(C)\right) + C
	\end{split}
	\label{eq:sample_average}
\end{align}
and its square
\begin{align}
	\begin{split}
	\EX\left[\left(\Delta G_{A \rightarrow B}^{(n)}\right)^2\right] &= \\- \int p_A(\x_1)\dx_1& ... \int p_A(\x_n)\dx_n \left[\ln \left( y^{(n)}(C)\right)+ C\right]^2 
	\label{eq:sample_average_square}
\end{split}
\end{align}  

are approximated by using the first order series expansion of the logarithm $\ln\left[y^{(n)}(C)\right] \approx y^{(n)}(C) - 1$ around $y^{(n)}(C) = 1$. Along similar lines, the exact difference and its square are approximated as $\Delta G_{A, B} = -\ln \left[y(C)\right] +C \approx -y(C)+1 + C$ and  $\left(\Delta G_{A, B}\right)^2 = \left(-\ln\left[ y(C)\right]+C \right)^2 \approx \left(-y(C)+1 + C \right)^2$ around $y(C) = 1$. 

Critically, for small $n$ the averages $y^{(n)}(C)$ and $y(C)$ generally differ, and therefore $C$ cannot be chosen such that both are approximately one. If, as in practice, $C$ is evaluated based on the acquired samples such that $y^{(n)}(C) = 1$, then $y(C)$ differs from one and, consequently, the first order series expansion of $y(C)$ becomes inaccurate. If $y^{(n)}(C)$ and $y(C)$ differ by, e.g., less than 10~\%, then the relative error of this approximation of the logarithm remains below 5~\%. However, for larger differences the neglected higher order terms will contribute markedly. A similar effect is caused by small configuration space density overlaps of the end states: Due to wider distributions of the exponentially weighted differences $H_B(\x)- H_A(\x)$, the variance of the sample based averages $y^{(n)}(C)$ will increase, and therefore also the average absolute deviations from  $y(C)$.
 
\begin{figure*}[t!]
	\includegraphics[width=\linewidth]{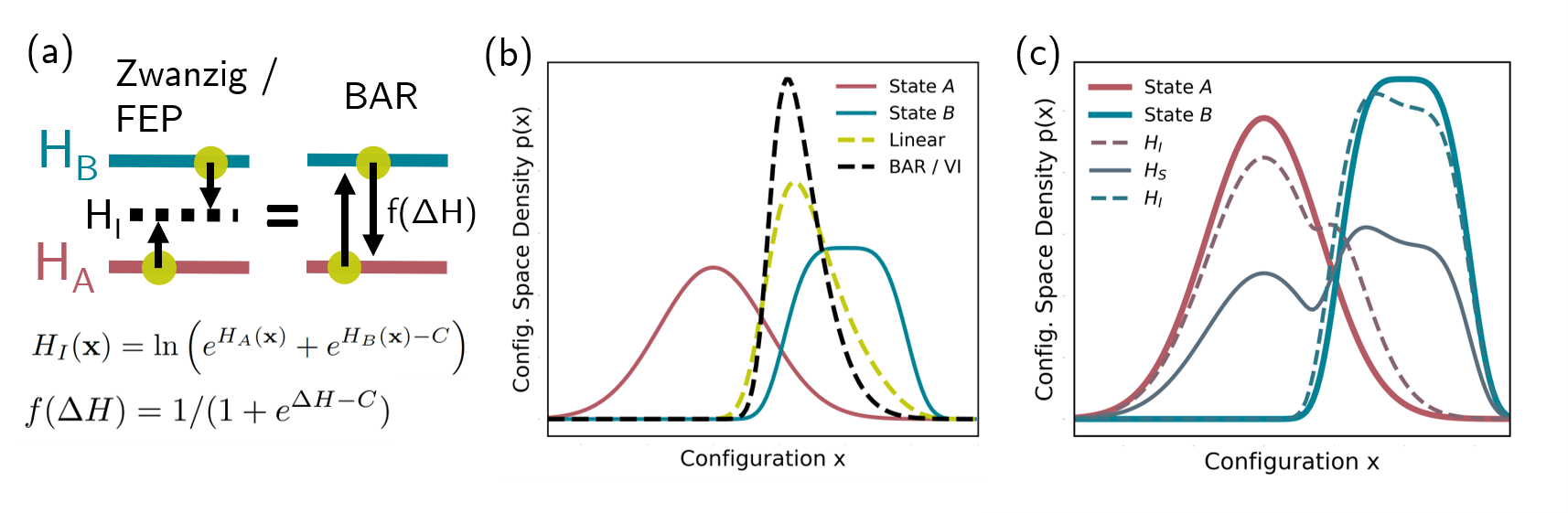}
	\caption{(a) Two schemes of free energy estimators. Left: Using the Zwanzig formula to calculate the free energy difference from the two end states to a virtual intermediate state in which no sampling is conducted. Right: Using BAR, where a weighting factor is applied to the difference in Hamiltonians. The two schemes are identical if the expressions shown beneath the schemes are used for the Hamiltonian of the virtual intermediate and the weighting function of BAR. (b) Configuration space densities of the virtual intermediate states corresponding to the linear estimator (green dashed line) and BAR (black dashed line). The densities of the harmonic end state, $H_A(\x) = ax^2$, and the quartic end state, $H_B(\x) = b(x-x_0)^4$, are shown in red and blue, respectively. (c) Variationally-derived Intermediates (VI). States in which sampling is conducted are indicated through solid lines, whereas virtual intermediates are indicated through dashed lines.}
	\label{fig:intro}
\end{figure*}

In the next step, Fig.~\ref{fig:intro}(a) shows how an intermediate state $I$ is used to derive the BAR formula via $ \Delta G_{A \rightleftharpoons B}^{(n)} = \Delta G_{A\rightarrow I}^{(n)} - \Delta G_{B\rightarrow I}^{(n)}$. We refer to $I$ as a {\em virtual} intermediate because it only serves as an end state for the Zwanzig formula, without being actually used for sampling. The derivation based on the above approximations~\cite{Reinhardt2020} yielded an additive MSE in this case, i.e., the MSE of the total estimate is 
\begin{align}
\mathrm{MSE}\left(\Delta G_{A\rightleftharpoons B}^{(n)}\right) = \mathrm{MSE}\left(\Delta G_{A\rightarrow I}^{(n)}\right) + \mathrm{MSE}\left(\Delta G_{B\rightarrow I}^{(n)}\right) \;.
\label{eq:mse_two_step}
\end{align}

For easier notation, we assume that the same number of samples $n$ is available for the two end states. Minimizing Eq.~(\ref{eq:mse_two_step}) through a variational approach leads to the Hamiltonian of the optimal virtual intermediate~\cite{Reinhardt2020} 
\begin{equation}
H_I(\x) = \ln \left(e^{H_A(\x)} + e^{H_B(\x)-C} \right)\;,
\label{eq:opt_virt_intermed}
\end{equation}
where the MSE is minimal if $C= \Delta G_{A,B}$ and approaches that minimum as $C$ approaches $\Delta G_{A,B}$. Figure~\ref{fig:intro}(b) shows this virtual intermediate state as a black dashed line for a one-dimensional example where one of the two end Hamiltonians is harmonic (red), and the other quartic (blue).  

Let us compare the result using $ \Delta G_{A \rightleftharpoons B}^{(n)} = \Delta G_{A\rightarrow I}^{(n)} - \Delta G_{B\rightarrow I}^{(n)}$ with intermediate Eq.~(\ref{eq:opt_virt_intermed}) to the original approach by Bennett~\cite{Bennett1976},  
\begin{align}
	\Delta G^{(n)}_{A \rightleftharpoons B} 
	= \ln \frac{\langle w(H_{A}(\x), H_{B}(\x)) e^{-H_{A}(\x)}\rangle_{B}}{\langle w(H_{A}(\x), H_{B}(\x)) e^{-H_{B}(\x)}\rangle_{A}} \, .
	\label{eq:bar_weighting}
\end{align}
which uses a suitably chosen weight function $w(H_{A}(\x), H_{B}(\x))$. Bennett optimized the weighting function with respect to the variance, which yields the widely used BAR result
\begin{align}
	\Delta G^{(n)}_{A, B} - C= \ln \frac{\langle f(H_{A}(\x) - H_{B}(\x) - C)\rangle_{B}}{\langle f(H_{B}(\x) - H_{A}(\x) + C)\rangle_{A}} \, ,
	\label{eq:zwanzig_two_step}
\end{align}
where $f(x) = 1\slash(1+e^x)$ is the Fermi function and $C\approx \Delta G_{A, B}$ has to be determined iteratively. 

From Eq.~(\ref{eq:bar_weighting}) and $ \Delta G_{A \rightleftharpoons B}^{(n)} = \Delta G_{A\rightarrow I}^{(n)} - \Delta G_{B\rightarrow I}^{(n)}$ with Eq.~(\ref{eq:zwanzig})  follows that the two approaches are equivalent if the weighting function relates to the Hamiltonian of the virtual intermediate state through 
\begin{align}
	w(H_{A}(\x), H_{B}(\x)) = e^{-H_I(\x) + H_{A}(\x) + H_{B}(\x)} \,.
	\label{eq:equivalence}
\end{align}

Therefore, any Hamiltonian of a virtual intermediate state corresponds to a weighting function. 

The variance of BAR~\cite{Bennett1976} is given by
    \begin{align}
   &\mathrm{Var}\left(\Delta G_{A,B}^{(n)}\right) = \frac{2}{n}\left[ \Omega^{-1} -1\right] \;\;, \label{eq:mse_bennett}\\
   \Omega = & \int \,\dx\, \frac{2p_A(\x)p_B(\x)}{p_A(\x) + p_B(\x)}
   \label{eq:omega}
   \end{align}
   where $\Omega$ can be interpreted as an overlap measure. Within the limits of the approximations discussed above, Bennett's variance, Eq.~(\ref{eq:mse_bennett}), equals the MSE, Eq.~(\ref{eq:mse_zwanzig}), of using Zwanzig in two steps, as is shown in Appendix A.

    This link between BAR and VI, Eq.~(\ref{eq:equivalence}), allows creating different estimators and transforming them between the formalism of using an intermediate state or a weighting function. Here, we will apply this result and compare BAR to the estimator that uses $H_I(\x) = \frac{1}{2}(H_A(\x)+ H_B(\x))$ as the virtual intermediate state. Because $H_I(\x)$  is a linear interpolation, we will refer to the resulting estimator as 'linear estimator', also known as the Simple Overlap Sampling method \cite{Lu2003, Lu2004}. The resulting configuration space density is shown by the green dashed line in Fig.~\ref{fig:intro}(b). As shown in Appendix B, our MSE for the Zwanzig formula, Eq.~(\ref{eq:mse_zwanzig}), yields the MSE for the linear estimator,
    \begin{align}
	\mathrm{MSE}\left(\Delta G_{A,B}^{(n)}\right) = \frac{2}{n}\left[\left(\int p_A(\x)^{\frac{1}{2}}p_B(\x)^{\frac{1}{2}}\dx\right)^{-2}-1\right] \;.
	\label{eq:mse_linear_interpolation}
    \end{align}
    The term in round brackets of Eq.~(\ref{eq:mse_linear_interpolation}) can be interpreted as an overlap measure, different from above, which equals one for two identical configuration space densities, and zero for disjunct supports. 

   Next, any number of optimal intermediate states can be derived by extending Eq.~(\ref{eq:mse_two_step}) with the MSEs of additional steps. Here, we focus our analysis on only one intermediate state $S$ for sampling, i.e., calculations of the form $A\rightarrow I\leftarrow S\rightarrow I \leftarrow B$. The optimization with variational calculus with respect to all intermediate Hamiltonians yields the VI. These consist of, firstly, Eq.~\ref{eq:opt_virt_intermed} (the BAR equivalent) as the optimal Hamiltonian of the virtual intermediates and secondly, the optimal sampling Hamiltonian $H_S(\x)$, which is determined through solution of 
    
    \begin{equation}
    \begin{split}
    H_S(\x)=-\frac{1}{2}\ln\Big[&\left(e^{H_A(\x)}\frac{Z_A}{Z_S} + e^{H_S(\x)}\right)^{-2}\\
    + &\left(e^{H_B(\x)}\frac{Z_B}{Z_S} + e^{H_S(\x)}\right)^{-2} \Big]\;\; .
    \end{split}
    \label{eq:soe_bar}
    \end{equation}
    The initially unknown ratios of the partition sums are determined iteratively, similar to the constant $C$ for BAR. The converged VI for the harmonic and quartic end states are shown in Fig.~\ref{fig:intro}(c).  
    
    In summary, for small $n$, BAR and VI result from the accurate optimization of an inaccurate MSE. Naturally, this does not ensure that better estimators and intermediate sampling states exist, which is therefore the subject of our test simulations.

    \section{Methods}
	
	In the first step, we assess the MSEs of different estimators. To this aim, we consider the one-dimensional system with end states consisting of a harmonic and a quartic Hamiltonian, as shown in Fig.~\ref{fig:intro}(b). Based on $n$ sample points drawn from the configuration space density of $A$ and $B$, the free energy estimate $\Delta G^{(n)}_{A \rightleftharpoons B}$ is obtained and compared to the exact difference $\Delta G_{A , B}$. Rejection sampling is used to obtain uncorrelated sample points. The MSE, Eq.~(\ref{eq:mse_ansatz}), is then calculated by averaging over one million of such realizations. We use $n=1, 20$ and 1000 sample points per end state. For each $n$, we consider 82 different setups for which the potential of end state $B$ is moved horizontally away from $A$ by varying $x_0$, thereby considering a range of overlap $\Omega$, which is obtained through numerical integration of Eq.~(\ref{eq:omega}). 
	
	With this procedure, we compare three variants. To separate the effects of an inaccurate estimate of $C$, firstly, BAR is used  where $C$ has been set to the (in practice unknown) exact free energy difference. Secondly, using BAR, where $C$ is iteratively determined based on the sample set as done in practice. Thirdly, the linear estimator. 
	
		\begin{figure*}[t!]
		\includegraphics[width=\linewidth]{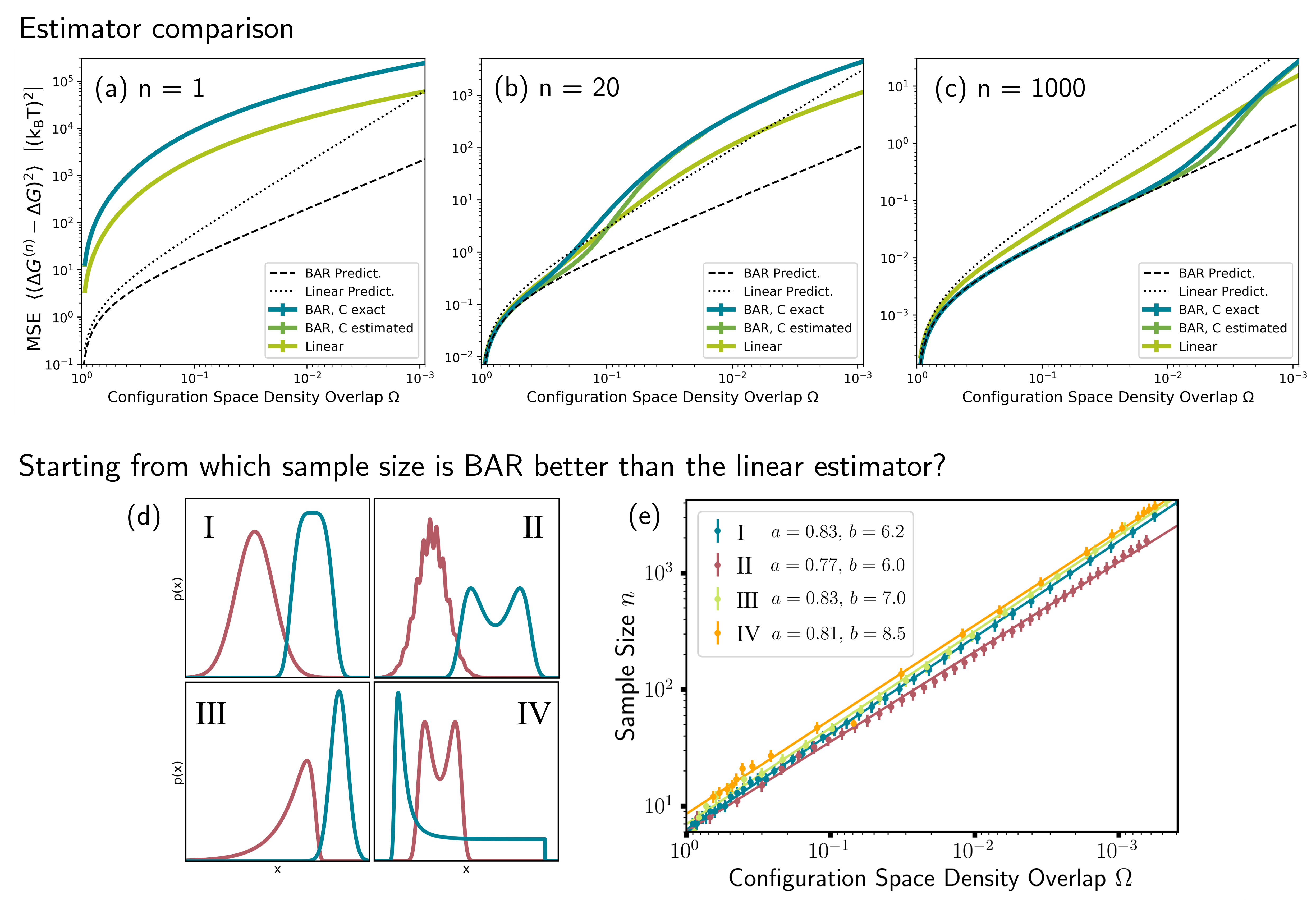}
		\caption{Comparison of BAR and the linear estimator. (a)-(c) MSEs obtained from test simulations based on the setup shown in Fig.~\ref{fig:intro}(b) for sample sizes of $n=1, 20$ and 1000. The MSEs are shown as a function of the configuration space density overlap $\Omega$, where different $\Omega$ were obtained by varying $x_0$ of the quartic end state. The results of two variants of BAR are shown: Firstly, using a constant $C$ that equals the exact free energy difference (blue), and secondly, for $C$ that was iteratively determined for each set of samples (green). The MSE of the linear estimator is shown in yellow. The dashed and the dotted lines show the analytical MSEs calculated based on approximations for BAR and the linear estimator, respectively, i.e. Eqs.~(\ref{eq:mse_bennett}) and~(\ref{eq:mse_linear_interpolation}). (d) Setups used for the test simulations yielding the results shown by the respective Roman numbers (e). Setup I is identical to the one in Fig.~\ref{fig:intro}(b). (e) The minimum sample size $n$ required such that BAR with an exact $C$ yields a better (i.e., smaller MSE) than the linear estimator is shown as a function of $\Omega$. The solid lines show the function $n=b\,\Omega^{-a}$ fitted to the data points in the respective colors. The fit coefficients $a$ and $b$ are provided in the legend.}
		\label{fig:estimator_panel}
	\end{figure*}

	In the second step, aside from sampling in the end states, sampling is also conducted in one intermediate state $S$ and a similar procedure as above is used to evaluate the MSEs of different Hamiltonians $H_S(\x)$. Separate sample sets in $S$ are used to evaluate the free energy differences to either end state, as using the same sample set would introduce correlations between the two step-wise free energy estimates that would require a different analytic approach as the one described above~\cite{Reinhardt2020a}. Again, three variants are compared: Firstly, the VI, i.e., Eqs.~(\ref{eq:opt_virt_intermed}) and~(\ref{eq:soe_bar}). For simplicity, only exact estimates for $C$ and the ratios of the partition sums are considered. Secondly, as a comparison, two variants with a linearly interpolated sampling Hamiltonian: One using the linear estimator, and another one using BAR to evaluate the step-wise free energy difference. Again, the procedure was conducted for $n=1, 20$ and 1000 sample points per sample set. 	
	
	\section{Results and Discussion}

	The MSEs of the three estimator variants are shown in Fig.~\ref{fig:estimator_panel}(a)-(c) for different configuration space density overlaps $\Omega$ between the harmonic and the quartic end state. The panels show this relation for different sample sizes $n$. As can be seen, for $n=1$ both variants of BAR (blue and green) are suboptimal for all $\Omega$, as they yield a larger MSE than the linear estimator (yellow). For $n=20$, it depends on $\Omega$ whether BAR is suboptimal. Here, a turning point exists, i.e., the linear estimator is only better for approximately $\Omega < 10^{-1}$, whereas both BAR variants yield better MSEs for larger $\Omega$. For $n=1000$, this turning point shifts towards smaller $\Omega$. Here, the BAR variants perform better for around $\Omega > 10^{-3}$. Note that as the end states are different in form, the largest achievable overlap is $\Omega = 0.935$ and therefore no MSE of zero can be seen in Fig.~\ref{fig:estimator_panel}(a)-(c), which would be expected for $\Omega = 1$. 
	
	Unexpectedly, whereas for most $n$ and $\Omega$ both BAR variants have very similar MSEs, the one in blue where ${C=\Delta G_{A,B}}$ (i.e., the exact free energy difference) was used yields slightly larger MSEs than the variant that uses a sample based estimate of $C$ (green). This finding is in contrast to the widespread belief that an estimation for $C$ that deviates from ${\Delta G_{A,B}}$ is a major contribution to the inaccuracy of BAR. 
    The reason for this behavior lies in the first order series expansions of $\ln y(C)$ and $\ln y^{(n)}(C)$, as shown in the context of Eqs.~(\ref{eq:sample_average}) and (\ref{eq:sample_average_square}) in the theory section. For small $n$, $y^{(n)}(C)$ and $y(C)$ differ, and $C$ can therefore not be chosen such that the requirement is met that both are close to one. As a consequence, even if ${C=\Delta G_{A,B}}$ such that $y(C)=1$, then the first order series expansion of $\ln y^{(n)}(C)$ becomes inaccurate, and the same holds true for the subsequent derivation of BAR. 
	
	The dashed lines in Fig.~\ref{fig:estimator_panel}(a)-(c) show the predicted MSEs for BAR, i.e., Eq.~(\ref{eq:mse_bennett}), whereas the dotted lines show the one of the linear estimator, Eq.~(\ref{eq:mse_linear_interpolation}). As can be seen from Fig.~\ref{fig:estimator_panel}(a), for $n=1$ the predicted MSEs are much too small. Furthermore, BAR is predicted to have a better MSE than the linear estimator which is not the case for the results of the test simulations. For $n=20$, the MSEs start to agree for large $\Omega$, but still deviate substantially for small $\Omega$. For BAR with $n=1000$, the MSEs agree well for most $\Omega$. For the linear estimator, the prediction is still mostly only accurate for large $\Omega$. Interestingly, unlike at $n=1$, Eq.~(\ref{eq:mse_linear_interpolation}) predicts an MSE that is larger than the one from the test simulations for $n=1000$. These results show that BAR is only optimal in cases where the predicted MSE is close to the actual one. In cases where the predicted MSE is inaccurate, BAR, as the optimization thereof, becomes suboptimal. 
	
	As the turning point $\Omega$ above which BAR becomes optimal varies with $n$, the question arises for the relation between the required $n$ for different $\Omega$ and how this relation compares for different systems. Therefore, in the next step we test how many sample points are required for BAR to achieve a smaller MSE than the linear estimator, depending on the configuration space density. To this aim, the first variant is used ($C$ exact). Starting with $n=1$, the MSEs of both BAR and the linear estimator are calculated and $n$ is gradually increased until the turning point is found. In addition to the setup consisting of end states with a harmonic and a quartic Hamiltonian, three other diverse systems are considered, which are shown in Fig.~\ref{fig:estimator_panel}(d). Again, for each system different horizontal shifts are used to vary $\Omega$. The definitions and parameters of these systems are described in Appendix~C.
	
	    	\begin{figure*}[t!]
		\includegraphics[width=\linewidth]{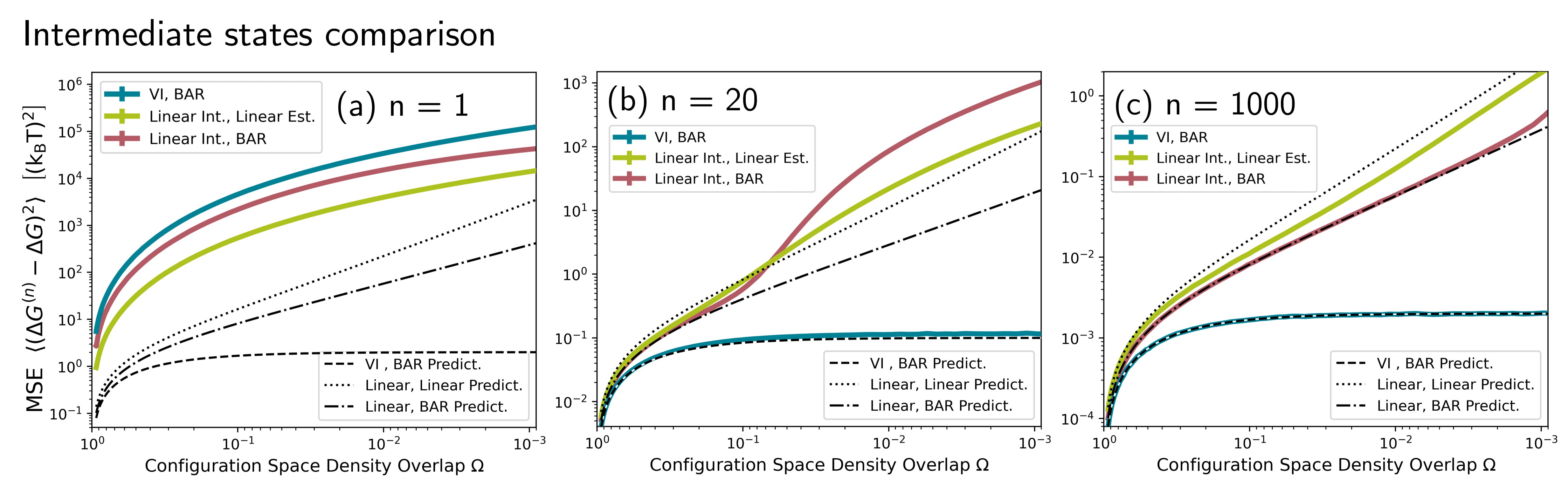}
		\caption{Comparison of the MSEs between using a linear intermediate state and VI. As for Fig.~\ref{fig:estimator_panel}(a)-(c), test simulations with a harmonic and a quartic end state were used, and (a)-(c) show the results for samples size of $n=1, 20$ and 1000, respectively, in each state as a function of the configuration space density overlap $\Omega$ between the end states. The results of two variants using a linear intermediate state are shown: Firstly, using the linear estimator (yellow) and secondly, using BAR (red) to evaluate the step-wise free energy differences. The MSE of VI, which includes using virtual intermediate states that correspond to BAR as shown in Fig.~\ref{fig:intro}(c) is shown in blue. The respective analytical MSEs are shown as black dashed, dotted and dashed-dotted lines. }
		\label{fig:intermediate_panel}
	\end{figure*}
	
    The required number of sample points $n$ is shown in dependence of $\Omega$ in Fig.~\ref{fig:estimator_panel}(e). The four colors indicate the different test systems with corresponding roman numbers from Fig.~\ref{fig:estimator_panel}(d). The required $n$ closely follow a linear relation in the log-log plot, indicating a relation of the form $n = b\,\Omega^{-a}$. Fits of this form are shown as solid lines and the fit coefficients are provided in the legend of Fig.~\ref{fig:estimator_panel}(e). Interestingly, the relation between $n$ and $\Omega$ is very similar for all four test systems, suggesting that $\Omega$ and $n$ are almost the sole factors that determine which estimator is superior.  
    
    Figure~\ref{fig:intermediate_panel}(a)-(c) compares MSEs for different intermediate sampling states $S$ as a function of the overlap $\Omega$ between $A$ and $B$ for $n=1, 20$ and 1000 per sample set. For $n=1$, the linear intermediate combined with the linear estimator (yellow) yields the best MSE, followed by the linear intermediate with BAR (red) and VI (blue) that includes BAR as an estimator. For $n=20$ and $n=1000$, VI yields the best MSE for all $\Omega$. For the linear intermediate sampling state, for $n=20$ a turning point exists ($\Omega \approx 5\cdot 10^{-2}$), above which BAR is superior, and below which the linear estimator is superior. For $n=1000$, BAR yields better MSEs at all $\Omega$. 
    	
    Again, for $n=1$ the predicted MSEs are much smaller than the actual ones. However, already for $n=20$, the actual MSE for VI is only slightly larger than the prediction, and matches perfectly for $n=1000$. For the linear intermediate, for $n=20$ both the predictions for BAR and the linear estimator hold only for larger overlaps. For $n=1000$, the one for BAR matches the actual MSEs very well, whereas for the linear estimator the prediction reproduces the trend but slightly overestimates the MSEs for small overlaps. We also tested how many sample points $n$ are required per state for VI to be optimal. Whereas for systems with large $\Omega$, two or three sample points per state suffice, in no case does the required number of sample points exceed seven per state (data therefore not shown). 
    
    These results show that, again, the predicted MSEs are inaccurate for small $n$. As a consequence, the VI, which have been derived as an optimization thereof, are suboptimal. However, using an intermediate sampling state, the MSEs become accurate and VI becomes optimal for much fewer $n$ than for BAR. We attribute this unexpected result mainly to the fact that for VI the sampling intermediate still maintains a large overlap with both end states, even if their configuration space densities are entirely disjunct. 

    \section{Summary and Conclusion}

    We have shown that for small sample sizes $n$ the analytically calculated MSEs of free energy estimates based on the Zwanzig formula become increasingly inaccurate due to approximations in its derivation. As a consequence, BAR and VI, which have been derived as an optimization thereof, become suboptimal for small $n$, which was demonstrated through the existence of better alternatives. For BAR, even if the constant $C$ is set to the exact free energy difference this suboptimality not only remains, but is even slightly worse than when $C$ is estimated based on the samples. 
   
    Whether BAR and VI are optimal depends, aside from $n$, on the configuration space density overlap $\Omega$, because for small $\Omega$ the fluctuations in the exponential averages increase. However, whereas BAR is suboptimal even for $n>1000$ if $\Omega < 10^{-3}$, VI is already better than all other tested variants for $n=7$ independent of $\Omega$, owing to the fact that the overlap between adjacent states is largely increased when using an intermediate state. For BAR, $\Omega$ was almost the sole factor that determined how many sample points were required to be better than the linear estimator. The relation follows an inverse power law of the form $n=a \Omega ^{-b}$, with very similar coefficients $a$ and $b$ for all four test systems considered. 
   
    It should be emphasized again that in atomistic simulations subsequent sample points are correlated, whereas the theory in this work relies on the common assumption of independent sample points. Therefore, the $n$ provided here such that BAR is optimal will typically refer to the {\em effective} number of statistically independent sample points, which is typically much smaller that the actual sample size. The low number effects on then MSE assessed here, therefore, can be relevant in macromolecular applications also for quite large sample sizes. 
    
    For such applications, instead of monitoring the variance or MSE directly (as implemented in many simulation software packages), we recommend to firstly consider $\Omega$. Secondly, packages such as alchemical-analysis.py \cite{Klimovich2015} analyze the time correlations between sample points and give an estimate for the number of independent ones. Then, thirdly, the relation between the required $n$ and $\Omega$ from this work will indicate whether BAR is optimal or whether another estimator such as the linear one should be used instead. Furthermore, in cases where BAR becomes close to being suboptimal, also the uncertainty estimates become inaccurate and other methods such as bootstrapping should be considered. Whereas BAR will remain the optimal estimator in many cases, these findings can help to assure that the optimal estimators are employed in all challenging applications.

    \FloatBarrier
	
	\appendix
	
	\section{Appendix A: Proof of MSE Equivalence to BAR Variance} 
	
	The Zwanzig formula~\cite{Zwanzig1954}, Eq.~(\ref{eq:zwanzig}), is used in two steps, as shown in Fig.~\ref{fig:intro}(a). The MSE of a single step is given through Eq.~(\ref{eq:mse_zwanzig}). Therefore, the total MSE is calculated through 
	
	\begin{align}
	&\;\;\;\;\mathrm{MSE}\left(\Delta G_{A\rightleftharpoons B}^{(n)}\right)\\ &= \mathrm{MSE}\left(\Delta G_{A\rightarrow I}^{(n)}\right) + \mathrm{MSE}\left(\Delta G_{B\rightarrow I}^{(n)}\right) \\
	&=  \frac{1}{n}\left(\int \left(p_I(\x)\right)^2 \left(\frac{1}{p_A(\x)} + \frac{1}{p_B(\x)}\right)\,\dx\,  - 2\right)\:.
	\label{eq:mse_zwanzig_two_step}
	\end{align}
    Using the configuration space density of the optimal virtual intermediate, Eq.~(\ref{eq:opt_virt_intermed}),
    \begin{align}
    p_I(\x) = \frac{\left[p_A(\x) ^{-1} + p_B(\x)^{-1}\right] ^{-1}}{\int\,\dx\,\left[p_A(\x) ^{-1} + p_B(\x)^{-1}\right] ^{-1}}
    \label{eq:mses_with_probs}
    \end{align}
	leads to
	\begin{align}
	\begin{split}
	&\mathrm{MSE}\left(\Delta G_{A\rightleftharpoons B}^{(n)}\right)\\ 
	=&\frac{1}{n}\frac{\int \,\dx\,\left[p_A(\x) ^{-1} + p_B(\x)^{-1}\right]^{-1}}{(\int\,\dx\,\left[p_A(\x) ^{-1} + p_B(\x)^{-1}\right] ^{-1})^2} -\frac{2}{n}\\
	=& \frac{1}{n}\left(\int\,\dx\,\frac{1}{p_A(\x) ^{-1} + p_B(\x)^{-1}}\right)^{-1} -\frac{2}{n}\\
	=& \frac{1}{n}\left(\int\,\dx\,\frac{p_A(\x)p_B(\x)}{p_A(\x) + p_B(\x)}\right)^{-1} -\frac{2}{n} \;,
	\end{split}
	\end{align}
    which equals the variance from Bennett~\cite{Bennett1976}, Eq.~(\ref{eq:mse_bennett}). 
	
	\section{Appendix B: MSE Derivation of the Linear Estimator}
	
    The linear estimator uses the linear interpolation $H_I(\x)= \frac{1}{2}\left(H_A(\x) + H_B(\x)\right)$ as the virtual Hamiltonian. The corresponding MSE is calculated by inserting the configuration space density,
	\begin{align}
	p_I(\x) = \frac{e^{-\frac{1}{2}[H_A(\x) + H_B(\x)]}}{Z_I}
	\end{align}
	into the expression of the MSE for using Zwanzig in two steps, Eq.~(\ref{eq:mse_zwanzig_two_step}), which yields
	\begin{align}
	 &\mathrm{MSE_{lin}}\left(\Delta G_{A\rightleftharpoons B}^{(n)}\right) \nonumber\\
	 	\begin{split} 
	 	=& \frac{1}{n}\Bigg(\int \Bigg[\frac{e^{-[\Ha + \Hb]}}{\displaystyle \left(\int e^{-\frac{1}{2}[\Ha + \Hb]}\dx\right)^2} \\ & \left(\frac{Z_A}{e^{-\Ha}} + \frac{Z_B}{e^{-\Hb}}\right)\,\Bigg]\dx\,  - 2\Bigg) 
	 \end{split}\\
	 = & \frac{1}{n}\left(\frac{\displaystyle \int \left(Z_A e^{-\Hb} + Z_B e^{-\Ha}\right)\dx}{\displaystyle\left(\int e^{-\frac{1}{2}[\Ha + \Hb]}\dx\right)^2}-2\right) \\
	 = & \frac{1}{n}\left(\frac{2Z_A Z_B}{\displaystyle\left(\int e^{-\frac{1}{2}[\Ha + \Hb]}\dx\right)^2}-2\right) \\	 
	 = & \frac{2}{n}\left[\left(\int p_A(\x)^{\frac{1}{2}}p_B(\x)^{\frac{1}{2}}\dx\right)^{-2}-1\right] \;.
	\end{align}

    \section{Appendix C: Parameters of Test Systems}
    The test systems shown in Fig.~\ref{fig:estimator_panel}(d) are based on the Hamiltonians provided below. These were used to determine the results shown in Fig.~\ref{fig:estimator_panel}(e), i.e., the minimum required number of sample points $n$ as a function of $\Omega$ such that BAR yields a smaller MSE than the linear estimator. \\
    
   \noindent
System I: $H_A(\x) = 0.75 \,x^2$ and $H_B(\x) = (x-x_0)^4$ using 46 values for $x_0$ with  $ 0 \le x_0 \le 4.5 $. \\[0.5em]
System II: $H_A(\x) = 0.1\, \sin(20x) + x^2$ and $H_B(\x) = 0.3\,x^4 - 0.8\,(x-x_0)^2$ using 47 values for $x_0$ with $0\le x_0 \le 23$. \\[0.5em]
System III: $H_A(\x) = e^x - x$ and $H_B(\x) = 0.15\,(x-x_0)^2$ using 24 values for $x_0$ with $0\le x_0 \le 9$. \\[0.5em]
System IV: $H_A(\x) = 0.3\,x^4 - 0.8\,(x-x_0)^2$ and $H_B(\x) = 4\epsilon \left[\left(\frac{\sigma}{x-x_0}\right)^{12} - \left(\frac{\sigma}{x-x_0}\right)^6\right]$ for $0<x-x_0 \le15$ and $H_B(\x)=\infty$ otherwise, using $\epsilon = 2.0446$ and $\sigma = 3.405$ and 22 values for $x_0$ with $0\le x_0 \le 4.03$. 
    
    	\bibliography{smallsample}

\end{document}